\documentclass[12pt]{iopart}
\usepackage{iopams}
\usepackage{graphicx}
\usepackage{bm}
\usepackage[colorlinks=true,linkcolor=blue,citecolor=red]{hyperref}%
\usepackage{cite}
\usepackage{color}

\def\erfc{{\rm erfc}}
\def\erfcx{{\rm erfcx}}
\def\x{\bm{x}}

\begin{document}

\title{Survival in a partially reactive wedge}

\author{Denis S Grebenkov}

\address{Laboratoire de Physique de la Mati\`ere Condens\'ee,
CNRS -- Ecole Polytechnique, Institut Polytechnique de Paris,
91120 Palaiseau, France}

\ead{denis.grebenkov@polytechnique.edu}

\begin{abstract}
We investigate the power-law decay of the survival probability of a
Brownian particle diffusing in an infinite planar wedge whose two
sides are partially reactive and described by Robin boundary
conditions.  A formal matched asymptotic analysis predicts the
persistence exponent $\pi/(2\alpha)$ and expresses the long-time
prefactor in terms of a stationary harmonic Robin problem near the
apex.  For opening angles $\alpha = \pi/n$ with positive integer $n$,
we solve the stationary problem exactly and obtain an explicit
expression for the prefactor as a function of the dimensionless
starting position.  For other angles, we conjecture an explicit
formula for the apex prefactor and provide numerical evidence for it.
The stationary solution describes the crossover to the perfectly
absorbing wedge.  As immediate applications, we deduce the long-time
behavior of the probability density function of the boundary local
time on wedge sides, as well as the probability density function of
the associated first-crossing time.
\end{abstract}


\section{Introduction}

First-passage processes in domains with corners provide elementary
examples in which geometry strongly affects the long-time kinetics of
diffusion \cite{Redner2001,Bray2013,Grebenkov2005}.  A paradigmatic
example is Brownian motion in an infinite two-dimensional wedge of
opening angle $\alpha$.  For a particle started from a point $\x_0$
inside the wedge with perfectly absorbing sides, the survival
probability decays algebraically as
\begin{equation}
S_\infty(t|{\bm x}_0)\propto t^{-\pi/(2\alpha)}     \qquad (t\to\infty) ,
\end{equation}
with a persistence exponent $\pi/(2\alpha)$ that depends continuously
on the opening angle (for the exact form of this survival probability,
its properties and applications, see
\cite{Dy2008,Chupeau2015,LeVot2020,Krapivsky2022} and references therein).

The situation becomes substantially more difficult when the sides of
the wedge are only partially reactive.  Partial reactivity is
conventionally described by a Robin boundary condition, $\partial_n
S_q + q S_q = 0$, which introduces an intrinsic length scale $q^{-1}$.
In contrast to Dirichlet or Neumann boundary conditions, the Robin
condition does not separate in polar coordinates because the normal
derivative on a radial boundary contains a factor $1/r$.  This simple
geometrical feature makes the partially reactive wedge a nontrivial
problem despite the apparent simplicity of its geometry.  Recently,
well-posedness and regularity results for the heat equation with Robin
boundary conditions in an unbounded two-dimensional wedge were
established in \cite{Bravin2025} (see also \cite{Nursultanov2025} for
polygonal domains).  To our knowledge, however, the long-time
asymptotics of the survival probability, and in particular its
amplitude, have not been determined.

In Ref.~\cite{YeGrebenkov2025}, the statistics of the boundary local
time on wedge sides was investigated as a tool for describing
diffusion-controlled reactions on partially reactive polygonal
boundaries.  In particular, for a particle started at the apex, it was
argued that the survival probability should exhibit the long-time
behavior
\begin{equation}   \label{eq:oldconjecture}
S_q(t|0)\propto \left(q\sqrt{Dt}\right)^{-\pi/\alpha} ,
\end{equation}
where $D$ is the diffusion coefficient.  This prediction was based on
scaling arguments and comparison with the absorbing wedge, and was
stated as a conjecture.  An analytical derivation of
Eq. (\ref{eq:oldconjecture}), including the corresponding prefactor,
remained unavailable.

In this paper, we address this problem by matched asymptotic analysis.
The key observation is that the long-time dynamics involves two widely
separated length scales: the reaction scale $q^{-1}$ and the diffusion
scale $\sqrt{Dt}$.
On the diffusion scale $\sqrt{Dt}$, the Robin boundary behaves
asymptotically as an absorbing boundary.  In contrast, within a
distance of order $q^{-1}$ from the apex, partial reactivity remains
essential.  Matching these two regions reduces the determination of
the leading long-time asymptotics to a stationary harmonic Robin
problem in the wedge (see Sec. \ref{sec:main}) and thus yields the
long-time behavior
\begin{equation}   \label{eq:mainconjecture}
S_q(t|\x_0) \simeq B_\alpha(q\x_0) \left(q\sqrt{Dt}\right)^{-\pi/\alpha}  \,,
\end{equation}
with the space-dependent prefactor $B_\alpha(q\x_0)$.   Since the
exponent of the dimensionless variable $q\sqrt{Dt}$ is $\nu =
\pi/\alpha$, the decay is proportional to $t^{-\nu/2}$, so the
persistence exponent for partially reactive wedges is $\pi/(2\alpha)$.
Most importantly, we go beyond the exponent analysis
\cite{Levernier2019} and determine explicitly the prefactor
$B_\alpha(q\x_0)$ for wedges of angle $\alpha = \pi/n$ with integer
$n$.  We show therefore how partial reactivity and starting point
affect the survival probability at long times.  We further conjecture
a simple explicit form of $B_\alpha(0)$ for arbitrary opening angles
and provide numerical evidence in its support.  In
Sec. \ref{sec:conclusion}, we discuss two immediate applications of
this asymptotic result and conclude the paper.

\section{Survival probability}
\label{sec:main}

We consider the infinite wedge of angle $0 < \alpha < 2\pi$, defined
in polar coordinates $(r,\phi)$ as
\begin{equation*}
 \Omega_\alpha=\{(r,\phi):r>0,\; -\alpha/2 <\phi<\alpha/2\}.
\end{equation*}
For a starting point $\x_0 = (r,\phi)\in\Omega_\alpha$, the survival
probability $S_q(t|\x_0)$ satisfies the diffusion equation
\begin{equation*} 
\frac{\partial S_q}{\partial t} = D\left(\frac{\partial^2 S_q}{\partial r^2}
 +\frac{1}{r}\frac{\partial S_q}{\partial r}  +\frac{1}{r^2}\frac{\partial^2S_q}{\partial\phi^2}\right)   
\qquad \textrm{in}~~\Omega_\alpha,
\end{equation*}
with the initial condition $S_q(0|\x_0)=1$.  On the two sides of the
wedge we impose the Robin boundary condition,
\begin{equation*}  
\partial_n S_q + q S_q=0,
\end{equation*}
where $\partial_n$ denotes the outward normal derivative and $q>0$ has
dimensions of inverse length.

\subsection{Outer asymptotic solution}

We first recall the behavior of the survival probability in a
perfectly absorbing wedge (the limit $q = \infty$).  In the long-time
regime $r/\sqrt{Dt}\ll 1$, the leading contribution comes from the
lowest angular eigenmode \cite{Redner2001,Dy2008,Chupeau2015}
\begin{equation}   \label{eq:outer}
S_\infty(t|\x_0) \simeq A_\nu  \left(\frac{r}{\sqrt{Dt}}\right)^\nu  \cos(\nu\phi),
\end{equation}
where
\begin{equation}    \label{eq:Anu}
A_\nu = \frac{2^{2-\nu} \, \Gamma(1+\nu/2)}{\pi \, \Gamma(1+\nu)}  \,, \qquad  \nu=\frac{\pi}{\alpha} \,.
\end{equation}

For finite $q$, the boundary condition can be rewritten schematically on
the diffusion scale as
\begin{equation*}
\frac{1}{\sqrt{Dt}}\partial_{\hat n}S_q + qS_q=0.
\end{equation*}
Thus, when $q\sqrt{Dt}\gg1$, the leading outer boundary condition is
Dirichlet, and Eq. (\ref{eq:outer}) therefore also provides the
leading outer solution of the partially reactive problem in the
overlap region $q^{-1}\ll r\ll\sqrt{Dt}$.

\subsection{Inner Robin problem}

On the inner scale, we hold $\rho = qr$ fixed and introduce the
dimensionless time $\tau = Dq^2t$.  For an inner term with a power-law
amplitude in $\tau$, differentiation in time introduces an extra
factor $\tau^{-1}$, whereas the spatial derivatives are taken at fixed
inner scale.  Thus the leading inner solution is quasistationary, and
we seek
\begin{equation}   \label{eq:inneransatz}
S_q(t|\x_0)\simeq K(t) \, u_\nu(\rho,\phi),
\end{equation}
where
\begin{equation}   \label{eq:harmonic}
\Delta u_\nu =0  \qquad \textrm{in}~~ \Omega_\alpha,
\end{equation}
with parameter-free Robin boundary conditions:
\begin{equation}  \label{eq:innerRobin1}
\pm \frac{1}{\rho}\partial_\phi u_\nu + u_\nu =0  \qquad \textrm{on}~~ \phi= \pm \frac{\alpha}{2} \,.
\end{equation}
By symmetry, $u_\nu(\rho,\phi) = u_\nu(\rho,-\phi)$, and we choose the
normalization $u_\nu(0,0)=1$.

Matching the inner solution (\ref{eq:inneransatz}) to the lowest
Dirichlet angular mode from Eq. (\ref{eq:outer}) requires
\begin{equation}   \label{eq:bdef}   
u_\nu(\rho,\phi) = b_\nu\, \rho^\nu \cos(\nu\phi) + o(\rho^\nu)  \qquad (\rho\to\infty),
\end{equation}
uniformly for $|\phi|\le\alpha/2$, where $b_\nu$ is a dimensionless
connection coefficient.  The apex normalization $u_\nu(0,0)=1$
fixes the far-field amplitude: $b_\nu =
\lim\nolimits_{\rho\to\infty}\rho^{-\nu}u_\nu(\rho,0)$.
Substitution of Eq. (\ref{eq:bdef}) into Eq. (\ref{eq:inneransatz})
and matching with Eq. (\ref{eq:outer}) determines $K(t)$ via
\begin{equation*}
K(t) b_\nu q^\nu = A_\nu (Dt)^{-\nu/2} \,.
\end{equation*}
Substituting this $K(t)$ into Eq. (\ref{eq:inneransatz}) yields the
asymptotic relation (\ref{eq:mainconjecture}), with
\begin{equation} \label{eq:Balpha}
B_\alpha(q\x_0) = \frac{A_\nu}{b_\nu}\, u_\nu(qr,\phi),
\end{equation}
where $A_\nu$ is given in Eq. (\ref{eq:Anu}).  The matching argument
therefore yields the persistence exponent $\pi/(2\alpha)$ for any
$\alpha$, whereas the computation of the spatial prefactor is reduced
to finding the solution $u_\nu$ of the stationary Robin problem
(\ref{eq:harmonic}, \ref{eq:innerRobin1}, \ref{eq:bdef}).  The
matching result (\ref{eq:Balpha}) does not assume any formula for the
coefficient $b_\nu$.  Moreover, at the apex, the normalization
$u_\nu(0,0) = 1$ gives $B_\alpha(0) = A_\nu/b_\nu$, so that only
$b_\nu$ remains to be determined.  In the next subsection, we compute
$b_\nu$ exactly for positive integer $\nu$, whereas its conjectural
extension for any $\nu > 0$ is introduced in Sec. \ref{sec:any_angle}.

\subsection{Exact solution for $\alpha=\pi/n$}

We now consider
\begin{equation*}
\alpha = \frac{\pi}{n},  \qquad n=1,2,\ldots,
\end{equation*}
so that $\nu=n$.  A regular harmonic function symmetric with respect
to the wedge bisector can be expanded as
\begin{equation*}
u_n(\rho,\phi) = \sum_{m=0}^{\infty} c_m \rho^m \cos(m\phi),  \qquad c_0=1.
\end{equation*}
Applying the boundary condition (\ref{eq:innerRobin1}) at
$\phi=\alpha/2$ gives
\begin{equation*}
-(m+1)c_{m+1} \sin\left(\frac{(m+1)\alpha}{2}\right) +  c_m\cos\left(\frac{m\alpha}{2}\right)=0  \qquad (m=0,1,2,\ldots),
\end{equation*}
and hence
\begin{equation*}
c_{m+1} =  c_m \frac{\cos(m\alpha/2)} {(m+1)\sin[(m+1)\alpha/2]} \,,
\end{equation*}
from which
\begin{equation}  \label{eq:am}
c_m = \frac{1}{m!} ~ \frac{\displaystyle\prod_{j=0}^{m-1} \cos\left(\frac{j\pi}{2n}\right)}
 {\displaystyle\prod_{j=1}^{m} \sin\left(\frac{j\pi}{2n}\right)} \qquad (1\leq m \leq n).
\end{equation}
Setting $m = n$ in Eq. (\ref{eq:am}), changing the product index $j\to
n-j$ in the numerator and using $\cos(\pi(n-j)/(2n)) = \sin(\pi
j/(2n))$, one readily finds that the numerator and denominator
products cancel, yielding
\begin{equation*}
c_n = \frac{1}{n!} \,.
\end{equation*}
Moreover, since $\cos(n\alpha/2) = \cos(\pi/2)=0$, one has $c_{n+1}=0$
and all subsequent coefficients vanish.  The inner solution is
therefore an explicit harmonic polynomial of degree $n$:
\begin{equation}    \label{eq:polynomial}
u_n(\rho,\phi) = \sum_{m=0}^{n} c_m \rho^m \cos(m\phi).
\end{equation}
Comparing its leading large-$\rho$ behavior, $u_n(\rho,\phi) \simeq
\rho^n \cos(n\phi)/n!$, with Eq. (\ref{eq:bdef}) gives the exact
connection coefficient
\begin{equation}   \label{eq:bn}
b_n = \frac{1}{n!}\,.
\end{equation}
Substituting Eqs. (\ref{eq:Anu}) and (\ref{eq:bn}) into
Eq. (\ref{eq:Balpha}), we obtain our main result
(\ref{eq:mainconjecture}) for $\alpha = \pi/n$, with
\begin{equation} \label{eq:Balpha2}
B_\alpha(q\x_0) = \frac{2^{2-n}}{\pi} \Gamma\left(1+\frac{n}{2}\right) u_n(qr,\phi).
\end{equation}

When the starting point $\x_0$ is at the apex, we get 
\begin{equation}    \label{eq:maininteger0}
S_q(t|0) \simeq \frac{2^{2-n}}{\pi} \Gamma\left(1+\frac{n}{2}\right) \left(q\sqrt{Dt}\right)^{-n} 
\qquad \left(q\sqrt{Dt}\gg 1\right).
\end{equation}
As a check, Eq. (\ref{eq:maininteger0}) reproduces the explicitly
known solutions for the half-plane ($\alpha=\pi$) and for the quadrant
($\alpha =\pi/2$).  For $n = 1$, the problem is effectively
one-dimensional and one finds $S_q(t|0) = \erfcx(q\sqrt{Dt})$, where
$\erfcx(z) = e^{z^2} \erfc(z)$ is the scaled complementary error
function.  Using the large-argument expansion of the complementary
error function, one has $S_q(t|0) \simeq 1/(q\sqrt{\pi Dt})$ as $t\to
\infty$, in agreement with Eq. (\ref{eq:maininteger0}).  For $n = 2$,
the two Cartesian coordinates are independent so that $S_q(t|0) =
[\erfcx(q\sqrt{Dt})]^2 \simeq 1/(\pi q^2Dt)$ as $t\to \infty$, again
in agreement with Eq. (\ref{eq:maininteger0}).

\subsection{Crossover to the absorbing-wedge asymptotics}

The general expression (\ref{eq:mainconjecture}) also provides a
useful interpretation of the role of the dimensionless distance $qr$
from the apex.  For $qr \ll 1$, the starting point lies within the
Robin corner region.  Since $u_\nu(0,0)=1$, the spatial prefactor
remains close to its apex value, with corrections determined by the
regular expansion of $u_\nu$.

In the opposite regime $q r \gg 1$, the starting point lies well
outside the characteristic Robin length $q^{-1}$.  Substitution of
Eq. (\ref{eq:bdef}) in Eqs. (\ref{eq:mainconjecture}) and
(\ref{eq:Balpha}) yields
\begin{equation*}
S_q(t|\x_0) \simeq A_\nu \left(\frac{r}{\sqrt{Dt}}\right)^\nu  \cos(\nu\phi).
\end{equation*}
The Robin parameter $q$ has disappeared from the leading term, and one
recovers precisely the long-time asymptotic behavior of the perfectly
absorbing wedge.  The inner harmonic function $u_\nu$ therefore
describes the crossover, as a function of $qr$, between the partially
reactive corner region and the effectively absorbing outer region.  We
stress that the limits considered here correspond to a fixed starting
point followed by $t\to\infty$.  More generally, if $r$ grows with
time so that $r/\sqrt{Dt}$ remains finite, the simple factorized
asymptotic form (\ref{eq:mainconjecture}) is no longer applicable and
the full outer solution has to be retained.

\subsection{Arbitrary angle}
\label{sec:any_angle}

The polynomial construction of the preceding section relies on $\nu =
\pi/\alpha$ being an integer.  Nevertheless, it suggests the natural
extension
\begin{equation}   \label{eq:bconjecture}
b_\nu = \frac{1}{\Gamma(1+\nu)} 
\end{equation}
for arbitrary $\nu>0$.  If Eq. (\ref{eq:bconjecture}) holds,
Eqs. (\ref{eq:Anu}) and (\ref{eq:Balpha}) immediately yield, for a
particle started at the apex:
\begin{equation} \label{eq:Balpha3}
B_\alpha(0) = \frac{2^{2-\pi/\alpha}}{\pi} \Gamma\left(1+\frac{\pi}{2\alpha}\right).
\end{equation}
For a generic starting point, however, an explicit determination of
the harmonic function $u_\nu(\rho,\phi)$, and hence of
$B_\alpha(q\x_0)$, remains an open problem.

In order to check Eq. (\ref{eq:bconjecture}) numerically, we consider
an auxiliary boundary value problem in a circular sector
$\Omega_{\alpha,R} = \{ (r,\phi) ~:~ 0 < \rho < R,~ |\phi|<
\alpha/2\}$ of angle $\alpha$ and of finite radius $R$:
\begin{equation*}
\Delta V_{\nu,R} =0  \qquad \textrm{in}~~ \Omega_{\alpha,R},
\end{equation*}
with Robin boundary conditions on the sides
\begin{equation*}
\pm \frac{1}{\rho}\partial_\phi V_{\nu,R} + V_{\nu,R} =0     \qquad \textrm{on}~~ \phi= \pm \frac{\alpha}{2} \,,
\end{equation*}
and Dirichlet boundary condition on the arc:
\begin{equation*}
V_{\nu,R}(R,\phi) = R^\nu \cos(\nu \phi).
\end{equation*}
By comparison in the conformally mapped half-plane of
Sec. \ref{sec:alternative}, $V_{\nu,R}$ converges locally uniformly to
the stationary profile $u_\nu/b_\nu$ with unit far-field coefficient;
in particular, $1/V_{\nu,R}(0,0)\to b_\nu$ as $R\to \infty$.

We solve the above boundary value problem by using the finite-element
solver of the MATLAB PDE Toolbox.  We used triangular elements of
polynomial degree $1$ and set the maximum mesh-size parameter to 0.05.
The reciprocals of the apex values, $1/V_{\nu,R}(0,0)$, computed at
five radii $R$ ranging from $10$ to $20$, were then extrapolated by
using a quadratic polynomial to estimate $b_\nu$.  We first validated
the numerical procedure for several values $\alpha = \pi/n$, for which
$b_n=1/n!$ is known exactly.  We then tested
Eq. (\ref{eq:bconjecture}) for several non-integer values of $\nu =
\pi/\alpha$.  Table \ref{tab:example} shows excellent agreement
between the numerical values of $b_\nu$ and its theoretical prediction
in Eq. (\ref{eq:bconjecture}).  The integer values $\nu = 2$ and $\nu
= 3$ are shown for an independent validation of this numerical
approach.

\begin{table}
\begin{center}
\begin{tabular}{ c | c | c | c | c | c | c | c }
$\nu~\backslash~ R$  & 10     & 12.5   &   15   & 17.5   & 20     & $\infty$ &  conj. $b_\nu$  \\ \hline
$2/3$                & 1.2196 & 1.1975 & 1.1827 & 1.1721 & 1.1641 & 1.1080   & 1.1077 \\
$4/3$                & 1.0035 & 0.9691 & 0.9466 & 0.9307 & 0.9190 & 0.8393   & 0.8399 \\
$3/2$                & 0.9281 & 0.8909 & 0.8666 & 0.8495 & 0.8369 & 0.7518   & 0.7523 \\
$2$                  & 0.6930 & 0.6510 & 0.6239 & 0.6051 & 0.5912 & 0.4998   & 0.5    \\
$5/2$                & 0.4816 & 0.4408 & 0.4149 & 0.3971 & 0.3840 & 0.3012   & 0.3009 \\
$3$                  & 0.3158 & 0.2805 & 0.2585 & 0.2436 & 0.2328 & 0.1674   & 0.1667 \\
\end{tabular}
\end{center}
\caption{
Numerical values of $1/V_{\nu,R}(0,0)$ at five truncation radii $R$
and several $\nu$, obtained by a FEM method, with the maximum mesh
size of $0.05$.  The next-to-last value is obtained by quadratic
extrapolation of $1/V_{\nu,R}(0,0)$ versus $1/R$ over these five
values.  The last column gives the conjectured value $b_\nu =
1/\Gamma(1+\nu)$.  Two values $\nu = 2$ and $\nu =3$, for which the
relation $b_\nu = 1/\Gamma(1+\nu)$ is rigorously established, are
included for an independent validation of the numerical approach.}
\label{tab:example}
%
\end{table}

\subsection{Alternative formulation of the connection problem}
\label{sec:alternative}

It is instructive to note that the coefficient $b_\nu$ can also be
characterized by a one-dimensional fractional boundary-value problem.
For this purpose, we derive an alternative representation of the inner
problem via the conformal mapping
\begin{equation}
w = z^\nu,  \qquad    z = \rho \, e^{i\phi},  \qquad  \nu = \frac{\pi}{\alpha} \,,
\end{equation}
which maps the wedge $|\phi|<\alpha/2$ onto the right half-plane
$\mathrm{Re}(w)>0$.  Writing $w=x+iy$ and $H(x,y) = u_\nu(z(w))$, we
note that $H$ is harmonic in the half-plane.  Since
$|dw/dz|=\nu\rho^{\nu-1}$ and $|y|=\rho^\nu$ on its boundary, the
Robin condition transforms into
\begin{equation}
\partial_n H + \frac{1}{\nu}|y|^{1/\nu-1} H=0     \qquad \textrm{on}~~ x=0.
\end{equation}

Under the conformal map, Eq. (\ref{eq:bdef}) becomes $H(x,y) = b_\nu x
+ W(x,y)$, where $W$ is harmonic and sublinear at infinity.  For the
stationary profile constructed by comparison, $W = O(|w|^{\mu/\nu})$
as $|w|\to\infty$, for some $\mu<\nu$.  Let $g(y) = W(0,y) = H(0,y)$
be its boundary trace.  Using the standard realization of $|D_y|=
(-\partial_y^2)^{1/2}$ as the Dirichlet-to-Neumann operator for
harmonic extension to a half-plane \cite{Caffarelli2007}, one has
$\partial_n W = |D_y|g$ and thus obtains the one-dimensional
fractional equation
\begin{equation}    \label{eq:fractional-g}
\left[|D_y|+\frac{1}{\nu}|y|^{1/\nu-1}\right]g(y) = b_\nu,    \qquad g(0)=1 .
\end{equation}
Equation (\ref{eq:fractional-g}), together with the prescribed growth
of the trace, gives an equivalent characterization of $b_\nu$.  For
integer $\nu = n$, Eq. (\ref{eq:fractional-g}) admits a terminating
fractional-power expansion, from which $b_n = 1/n!$ can be recovered
independently (not shown).  An explicit evaluation from this
formulation for noninteger $\nu$ remains open.

Note that if $|D_y|g(y) = o(1)$ as $|y|\to\infty$,
Eq. (\ref{eq:fractional-g}) yields the expected trace asymptotic
\begin{equation}    \label{eq:g-large-y}
g(y)\simeq \nu b_\nu |y|^{1-1/\nu}.
\end{equation}
In the original wedge, this gives the leading boundary value
$u_\nu(\rho,\pm\alpha/2)\simeq \nu b_\nu\rho^{\nu-1}$ as $\rho \to
\infty$, where the leading Dirichlet mode in Eq. (\ref{eq:bdef})
vanishes.

\section{Discussion and Conclusion}
\label{sec:conclusion}

We have investigated the survival probability of a Brownian particle
in a partially reactive wedge at long times.  The presence of a finite
Robin parameter introduces the microscopic length $q^{-1}$, whereas
diffusion generates the growing scale $\sqrt{Dt}$.  Their separation
at long times naturally leads to a matched asymptotic description.
The outer problem is governed by an effectively absorbing wedge and
therefore fixes the persistence exponent $\pi/(2\alpha)$.  All
information about finite boundary reactivity and starting point is
contained in the stationary inner problem at distances $r=O(q^{-1})$
from the apex.  In this sense, the long-time prefactor is a connection
coefficient between the Robin boundary layer near the corner and the
lowest Dirichlet mode at large distances, and thereby describes the
crossover between the Robin and Dirichlet regimes.

For the infinite family of wedges of angle $\alpha=\pi/n$ with
positive integer $n$, we have solved this connection problem exactly.
The harmonic expansion terminates after $n$ terms and produces a
polynomial whose highest-order coefficient is simply $b_n = 1/n!$.
This leads to the explicit form (\ref{eq:Balpha2}) for the prefactor
$B_\alpha(\x_0)$ in front of the asymptotic relation
(\ref{eq:mainconjecture}).  The integer result suggests the natural
extension (\ref{eq:bconjecture}) to an arbitrary wedge angle $\alpha =
\pi/\nu$.  Numerical solutions of the stationary Robin problem for
several non-integer values of $\nu$ strongly support this conjecture.
The coefficient $b_\nu$ can alternatively be characterized through the
one-dimensional nonlocal equation (\ref{eq:fractional-g}).
Establishing the conjectural extension (\ref{eq:bconjecture})
rigorously remains an interesting mathematical problem.  We also
stress that our matching analysis determines the leading profile and
its amplitude.  A rigorous justification of
Eq. (\ref{eq:mainconjecture}) would require control of the parabolic
remainder on the scale of the leading term, in particular in the
overlap $q^{-1}\ll r\ll\sqrt{Dt}$, where the outer Dirichlet profile
tends to zero.  More formally, one can formulate

{\bf Open mathematical question:} For fixed $0< \alpha < 2\pi$, fixed
$q>0$ and fixed $\x_0 = (r,\phi)\in \overline{\Omega}_\alpha$,
prove that
\begin{equation*}
\fl \qquad
\lim\limits_{t\to\infty} \bigl(q\sqrt{Dt}\bigr)^\nu S_q(t|\x_0) = \frac{A_\nu}{b_\nu}\,u_\nu(qr,\phi),
\qquad \nu = \frac{\pi}{\alpha} \,, \quad A_\nu = \frac{2^{2-\nu} \, \Gamma(1+\nu/2)}{\pi \, \Gamma(1+\nu)} \,,
\end{equation*}
where $u_\nu$ is the apex-normalized stationary Robin profile
satisfying (\ref{eq:harmonic}, \ref{eq:innerRobin1}, \ref{eq:bdef}),
and $b_\nu$ is its far-field coefficient.  The question concerns
convergence of the rescaled survival probability and is independent of
the conjectured explicit value of $b_\nu$ in
Eq. (\ref{eq:bconjecture}).

The asymptotic behavior (\ref{eq:mainconjecture}) has two immediate
implications.  As established in \cite{Grebenkov2020} and further
explored in \cite{YeGrebenkov2025}, the survival probability is the
generating function of the boundary local time $\ell_t$, which
quantifies the accumulated contact of the Brownian particle with the
boundary (see \cite{YeGrebenkov2025,Grebenkov2020} for its definition
and properties).  In fact, one has
\begin{equation} \label{eq:Sq_rho}
S_q(t|\x_0) = {\mathbb E}_{\x_0}\{ e^{-q\ell_t}\} = \int\limits_0^\infty d\ell \, e^{-q\ell} \, \rho(\ell,t|\x_0),
\end{equation}
where $\rho(\ell,t|\x_0)$ is the probability density function (PDF) of
$\ell_t$ when the particle started from $\x_0$.  A formal inversion of
the leading large-$q\sqrt{Dt}$ expression in the left-hand side of
Eq. (\ref{eq:Sq_rho}) gives, for a start at the apex,
\begin{equation}  \label{eq:rho_asympt}
\rho(\ell,t|0) \simeq \frac{B_\alpha(0)}{\Gamma(\pi/\alpha)} \, \frac{\left(\ell/\sqrt{Dt}\right)^{\frac{\pi}{\alpha} -1}}{\sqrt{Dt}} 
\qquad  \left(\ell/\sqrt{Dt} \ll 1\right),
\end{equation} 
with $B_\alpha(0) = A_{\pi/\alpha}/b_{\pi/\alpha}$, where $A_{\nu}$ is
given by Eq. (\ref{eq:Anu}) and $b_\nu$ is predicted by the
conjectural formula (\ref{eq:bconjecture}).  Since the small parameter
is $\ell/\sqrt{Dt}$, this relation also determines the small-$\ell$
behavior of the PDF.  Moreover, if $\alpha = \pi/n$, one can use the
explicit form (\ref{eq:Balpha2}) to extend this result to arbitrary
fixed starting points $\x_0 \ne 0$.  In fact, the explicit polynomial
dependence (\ref{eq:polynomial}) of $u_n(qr,\phi)$ on $q$ allows one
to formally invert the leading Laplace-transform expression in
Eq. (\ref{eq:Sq_rho}) term by term:
\begin{equation}  \label{eq:rho_asympt_x0}
\fl
\rho(\ell,t|\x_0) \simeq \frac{2^{2-n}}{\pi}\, \frac{\Gamma(1+n/2)}{(Dt)^{n/2}} 
\left(\frac{r^n \cos(n\phi)}{n!} \delta(\ell) + \ell^{n-1} \sum\limits_{m=0}^{n-1} c_m \cos(m\phi) \, \frac{(r/\ell)^m}{\Gamma(n-m)}\right),
\end{equation} 
with $c_m$ given by Eq. (\ref{eq:am}).  Here we used that the inverse
Laplace transform of $q^{m-n}$ is $\ell^{n-m-1}/\Gamma(n-m)$ for
$m<n$, and $\delta(\ell)$ for $m=n$.
The Dirac mass at $\ell=0$ represents trajectories that have avoided
both wedge sides up to time $t$.  Its exact weight is
$S_\infty(t|\x_0)$, whose leading asymptotic in Eq. (\ref{eq:outer})
gives the coefficient in front of $\delta(\ell)$ in
Eq. (\ref{eq:rho_asympt_x0}).  In turn, the regular part includes a
finite number of powers of $r/\ell$.

In the same vein, we also get access to the PDF of the first-crossing
time $T_\ell = \inf\{ t > 0 ~:~ \ell_t > \ell\}$, at which the
boundary local time first exceeds a prescribed threshold $\ell$.  In
fact, since $\ell_t$ is a nondecreasing process, one has
$\mathbb{P}_{\x_0}\{ \ell_t \leq \ell\} = \mathbb{P}_{\x_0}\{ T_\ell >
t\}$ for $\ell > 0$.  As a consequence, the PDF of $T_\ell$ is
\begin{equation}
U(\ell,t|\x_0) = - \partial_t \mathbb{P}_{\x_0}\{ T_\ell > t\} = - \partial_t \int\limits_0^{\ell} d\ell' \,
\rho(\ell',t|\x_0),
\end{equation}
where the integral includes the atom at zero.  Using the asymptotic
relation (\ref{eq:rho_asympt}), we obtain
\begin{equation}  \label{eq:U0_asympt}
U(\ell,t|0) \simeq \frac{B_\alpha(0)}{2\Gamma(\pi/\alpha)} \, \frac{\left(\ell/\sqrt{Dt}\right)^{\pi/\alpha}}{t} 
\qquad  \left(\ell/\sqrt{Dt} \ll 1\right).
\end{equation}
This asymptotic relation holds in the scaling regime
$\ell/\sqrt{Dt}\ll1$, and therefore describes in particular the
long-time behavior at fixed $\ell$ as well as the small-$\ell$
behavior.  When $\alpha = \pi/n$, we can use instead
Eq. (\ref{eq:rho_asympt_x0}) to get for $\ell > 0$
\begin{equation}  \label{eq:U_asympt_x0}
U(\ell,t|\x_0) \simeq \frac{n 2^{1-n}}{\pi}\, \frac{\Gamma(1+n/2)}{t} \left(\ell/\sqrt{Dt}\right)^n 
\sum\limits_{m=0}^{n} \frac{c_m \cos(m\phi)}{\Gamma(n-m+1)} \, (r/\ell)^m.
\end{equation} 
Equations (\ref{eq:U0_asympt}) and (\ref{eq:U_asympt_x0}) are the
formal predictions obtained by differentiating the leading cumulative
asymptotics in time.

Several extensions are worth mentioning.  First, it would be
interesting to determine subleading corrections to
Eq. (\ref{eq:mainconjecture}).  These corrections should contain
additional information about the crossover from the microscopic Robin
scale to the macroscopic Dirichlet regime.  Second, different
reactivities on the two sides of the wedge provide a natural
generalization in which both the inner connection problem and the
large-scale angular mode become asymmetric.  Third, the same matched
asymptotic mechanism should be relevant to polygonal domains, where
local corner geometries may generate long intermediate regimes
controlled by their wedge exponents.

More generally, the present analysis illustrates how partial
reactivity can regularize the singular starting point of a perfectly
absorbing wedge.  For Dirichlet boundary conditions, a particle
started exactly at the apex is absorbed immediately, whereas for any
finite $q$ its survival probability is nonzero and exhibits an
algebraic long-time decay.  The wedge angle determines the exponent,
while the Robin boundary layer fixes the amplitude.

\ack
The author used OpenAI's ChatGPT (GPT-5.6) during the preparation of
this manuscript to assist with language editing, manuscript
organization, and the exploration of analytical derivations.  The
author assumes full responsibility for all scientific content.


\section*{References}

\begin{thebibliography}{99}

\bibitem{Redner2001}		Redner S 2001
				{\it A Guide to First-Passage Processes}
				(Cambridge: Cambridge University Press)

\bibitem{Bray2013}		Bray A J, Majumdar S N and Schehr G 2013
				Persistence and first-passage properties in nonequilibrium systems
				{\it Adv. Phys.} {\bf 62} 225--361

\bibitem{Grebenkov2005}		Grebenkov D S 2005 
				What Makes a Boundary Less Accessible
				{\it Phys. Rev. Lett.} {\bf 95} 200602



\bibitem{Dy2008}		Dy D L L, Esguerra J P 2008
				First-passage-time distribution for diffusion through a planar wedge
				{\it Phys. Rev. E} {\bf 78} 062101  

\bibitem{Chupeau2015}		Chupeau M, B\'enichou O, Majumdar S N 2015
				Survival probability of a Brownian motion in a planar wedge of arbitrary angle
				{\it Phys. Rev. E} {\bf 91} 032106 

\bibitem{LeVot2020}		Le Vot F, Yuste S, Abad E, Grebenkov D S 2020
				First-encounter time of two diffusing particles in confinement
				{\it Phys. Rev. E} {\bf 102}, 032118

\bibitem{Krapivsky2022}		Krapivsky P L 2022
				Diffusion in a fluid flow generated by a source at the apex of a wedge
				{\it Phys. Rev. Fluids} {\bf 7} 064502 

\bibitem{Bravin2025}		Bravin M, Gnann M V, Kn\"upfer H, Masmoudi N, Roodenburg F B and Sauer J 2025
				Well-posedness and regularity of the heat equation with Robin boundary conditions in the two-dimensional wedge
				{\it Commun. Partial Differ. Equ.} {\bf 50} 1099--1134

\bibitem{Nursultanov2025}	Nursultanov M, Rowlett J, Sher D 2025
				The heat kernel on curvilinear polygonal domains in surfaces
				{\it Ann. Math. Qu\'ebec} {\bf 49} 1--61

\bibitem{YeGrebenkov2025}	Ye Y and Grebenkov D S 2025
				Boundary local time on wedges and prefractal curves
				{\it J. Phys. A: Math. Theor.} {\bf 58} 345002

\bibitem{Levernier2019}		Levernier N, Dolgushev M, B\'enichou O, Voituriez R, Gu\'erin T 2019
				Survival probability of stochastic processes beyond persistence exponents
				{\it Nat. Commun.} {\bf 10} 2990


\bibitem{Caffarelli2007}	Caffarelli L, Silvestre L 2007
				An extension problem related to the fractional Laplacian
				{\it Commun. Partial Differ. Equ.} {\bf 32} 1245--1260

\bibitem{Grebenkov2020}		Grebenkov D S 2020
				Paradigm Shift in Diffusion-Mediated Surface Phenomena 
				{\it Phys. Rev. Lett.} {\bf 125} 078102 

\end{thebibliography}
\end{document}